\documentclass[conference]{IEEEtran}
\IEEEoverridecommandlockouts
\usepackage{cite}
\usepackage{amsmath,amssymb,amsfonts}
\usepackage{algorithmic}
\usepackage{graphicx}
\usepackage{textcomp}
\usepackage{xcolor}
\usepackage{comment}
\usepackage{hyperref}
\usepackage{algorithm}
\usepackage{booktabs}
\usepackage{multirow}
\def\BibTeX{{\rm B\kern-.05em{\sc i\kern-.025em b}\kern-.08em
    T\kern-.1667em\lower.7ex\hbox{E}\kern-.125emX}}
\begin{document}

\title{Enhancing Efficiency and Performance in Deepfake Audio Detection through Neuron-level Dropin \& Neuroplasticity Mechanisms}
% \author{Anonymous submission}
% \begin{comment}

\author
{
\hspace*{-0.5cm}\IEEEauthorblockN{1\textsuperscript{st} Yupei Li\textsuperscript{*}}
\hspace*{-0.5cm}\IEEEauthorblockA{\hspace*{-0.5cm}\textit{Department of Computing and Chair of Health Informatics} \\
\hspace*{-0.5cm}\textit{Imperial College London and Technical University of Munich}\\
\hspace*{-0.5cm}London and Munich, United Kingdom and Germany \\
\hspace*{-0.5cm}yl7622@ic.ac.uk}
\and

\IEEEauthorblockN{2\textsuperscript{nd} Shuaijie Shao\textsuperscript{*}}
\IEEEauthorblockA{\textit{Department of Computer Science} \\
\textit{University College London}\\
London, United Kingdom \\
shuaijie.shao.25@ucl.ac.uk}
\and

% \IEEEauthorblockN{1\textsuperscript{st} Yupei Li\textsuperscript{*} \quad\hspace{5cm} 2\textsuperscript{nd} Shuaijie Shao\textsuperscript{*}}

% \IEEEauthorblockA{\textit{\hspace{-2.5cm}Department of Computing and Chair of Health Informatics} \hspace{2cm} \textit{Department of Computer Science} \\
% \textit{\hspace{-4cm}Imperial College London and Technical University of Munich} \quad\hspace{2cm}\textit{University College London} \\
% \hspace{-3cm}London and Munich, United Kingdom and Germany \hspace{2.5cm} London, United Kingdom \\
% yl7622@ic.ac.uk \quad\hspace{4cm} shuaijie.shao.25@ucl.ac.uk}
% \and

\IEEEauthorblockN{3\textsuperscript{rd} Manuel Milling}
\IEEEauthorblockA{\textit{Chair of Health Informatics} \\
\textit{Technical University of Munich}\\
Munich, Germany \\
manuel.milling@tum.de}
\and

\IEEEauthorblockN{4\textsuperscript{th} Bj\"orn Schuller}
\IEEEauthorblockA{\textit{Chair of Health Informatics and Department of Computing} \\
\textit{Technical University of Munich and Imperial College London}\\
Munich and London, Germany and United Kingdom \\
schuller@tum.de}
\and

\IEEEauthorblockA{
\hspace*{4cm}\textsuperscript{*} Yupei Li and Shuaijie Shao contributed equally to this work.
}
}
% \end{comment}
\maketitle
\begin{abstract}
Current audio deepfake detection has achieved remarkable performance using diverse deep learning architectures such as ResNet, and has seen further improvements with the introduction of large models (LMs) like Wav2Vec. The success of large language models (LLMs) further demonstrates the benefits of scaling model parameters, but also highlights one bottleneck where performance gains are constrained by parameter counts. Simply stacking additional layers, as done in current LLMs, is computationally expensive and requires full retraining. Furthermore, existing low-rank adaptation methods are primarily applied to attention-based architectures, which limits their scope. Inspired by the neuronal plasticity observed in mammalian brains, we propose novel algorithms, dropin and further plasticity, that dynamically adjust the number of neurons in certain layers to flexibly modulate model parameters. We evaluate these algorithms on multiple architectures, including ResNet, Gated Recurrent Neural Networks, and Wav2Vec. Experimental results using the widely recognised ASVSpoof2019 LA, PA, and FakeorReal dataset demonstrate consistent improvements in computational efficiency with the dropin approach and a maximum of around 39\% and 66\% relative reduction in Equal Error Rate with the dropin and plasticity approach among these dataset, respectively. The code and supplementary material are available at  \href{https://github.com/ShawnNotFound/Dropin}{Github link}.

\end{abstract}


\section{Introduction}
Large models (LMs) have recently surpassed the capability bottlenecks of traditional deep learning models across a wide range of tasks \cite{mckenzie2023inverse}, including audio-related applications such as deepfake audio detection. For instance, audio LLMs, such as QwenAudio \cite{chu2023qwen}, have demonstrated superior performance compared to conventional small-scale deep learning models. This performance improvement aligns with established scaling laws indicating increasing the number of model parameters generally enhances representational capacity and downstream task performance \cite{kaplan2020scaling} given sufficient training data and computation.

Additionally, we argue that large-scale models are beneficial for capturing the complexity of audio deepfake detection. Although this task is framed as a binary classification problem, the implicit discrepancies between spoofed and bona fide audio require the network to learn subtle and high-dimensional acoustic representations \cite{almutairi2022review}. Existing approaches employ diverse architectural paradigms for feature extraction: convolutional neural network (CNN)-based models such as ResNet \cite{he2015deepresiduallearningimage} process audio data as Mel-spectrograms by treating them as image-like features; recurrent neural network (RNN)-based architectures, including the Light Convolutional Gated Recurrent Neural Network (LC-GRNN), analyse spectrograms as temporally structured sequential data \cite{gomez2019light}; and attention-based pretrained models, notably Wav2Vec 2.0 \cite{baevski2020wav2vec20frameworkselfsupervised}, directly operate on raw audio waveforms to capture fine-grained temporal dependencies. Additional models have been proposed for this task, including ASSIST \cite{jung2022aasist} and RawNet2 \cite{tak2021endtoendantispoofingrawnet2}, etc. However, the approaches aforementioned have not fully resolved the challenges of audio deepfake detection, as evidenced by their performance in the recent ASVspoof 2019 challenge \cite{wang2020asvspoof2019largescalepublic}. Nevertheless, it is a consistent trend that models with more parameters tend to achieve superior performance.

These models commonly rely on fixed architectures, often incorporating fine-tuning model parameters pretrained on a different task.  
However, inspired by the mechanisms of the mammalian brain, it may be argued that modern models should aim for dynamic neuron-level adjustment called (neuro-)plasticity, enabling them to remain memory-efficient while unlocking additional capacity for target tasks. 
Li et al.\ \cite{li2025neuroplasticityartificialintelligence} reviewed the principles of plasticity in biology and deep learning methods that adopt similar dropin (adding more neurons) and plasticity concepts, yet without specific implementation details and any empirical tests. This idea has also been explored in continual learning, however, focus only on layer-level adjustments (i.e., adding or removing entire layers), such as Exhubert \cite{amiriparian2024exhubert}. In addition, many loss designs have been proposed to mitigate catastrophic forgetting \cite{dohare2024loss}, but these methods merely adjust weights that should be updated, without altering the number of parameters. 

Previous algorithms have only sparsely explored pioneering `drop-in–like' strategies, such as determining when and where to add neurons \cite{pmlr-v188-maile22a}. These methods identify triggers based on gradient information \cite{evci2022gradmax}, exposure to unseen experiences \cite{eriksson2019dynamic}, or model performance \cite{sowrirajan2024enhancing}. However, they do not explicitly draw inspiration from the mammalian brain to formulate dropin as a general technique; instead, their motivation is largely analytical, deriving from mathematical formulations aimed at increasing model capacity. Moreover, although various dropout and structural pruning strategies have been developed, they have not been integrated with dropin mechanisms to mirror the concepts of neuroplasticity.

Therefore, we propose a more fine-grained algorithm that incorporates a clearly defined dropin strategy alongside a novel, neuron-level plasticity mechanism. We evaluate this approach on multiple deepfake audio detection model clusters using the ASVspoof2019 LA, PA \cite{wang2020asvspoof2019largescalepublic}, and FakeorReal (FoR) \cite{reimao2019dataset} dataset. Experimental results show that the Equal Error Rate (EER) is considerably reduced and training time is shortened, all while maintaining limited memory usage. This leads to our \textbf{contributions}: To the best of our knowledge, this is the first neuron-level clearly defined dropin and novel plasticity algorithm that have been verified by experiments. Additionally, it enables state-of-the-art (SOTA) performance on ASVspoof2019 LA while preserving memory efficiency and reducing training time. Moreover, it is applicable to a variety of models.

\section{Related Work}
Previous work has explored brain-inspired methods from both biological and deep learning perspectives, providing evidence that such designs can improve both performance and computational efficiency \cite{wozniak2020deep, dellaferrera2021learning}. Representative examples include spiking neural networks with brain-like signalling mechanisms \cite{schmidgall2024brain}, Hebbian synaptic learning for representation learning \cite{ravichandran2025unsupervised, journe2022hebbian}, and hippocampus-inspired neuroplasticity algorithms \cite{rudroff2024neuroplasticity}. While these approaches demonstrate the potential of brain-inspired designs to enhance neural network performance, they are generally not scalable to a wide range of conventional neural network architectures.

Plasticity-related approaches, such as structural pruning and progressive networks, have been shown in prior work to be effective, although they are not always explicitly motivated by biological principles. Progressive networks \cite{rusu2016progressive} demonstrate that dynamically growing a network can enhance its learning capacity, while other studies increase model capacity by stacking multiple layers, for example, in ExHuBERT \cite{amiriparian2024exhubert} and NN-stacking \cite{coscrato2020nn}. Additionally, adaptive pruning methods based on feature information entropy have been proposed to reduce training costs while maintaining performance \cite{chen2025brain}. Moreover, synaptic plasticity–based regularisation techniques have been introduced to reduce the complexity of certain network architectures \cite{yousef2025synaptic}. However, these methods do not model a comprehensive process of neurogenesis and plasticity. 

Even works that more closely mimic neuroplasticity, such as determining neuron addition and pruning based on certain triggers, have been proposed. Related methods trigger insertion based on gradient information \cite{miconi2018differentiable}, exposure to novel experiences \cite{heidenreich2024transfer}, or other heuristics. However, these approaches are not grounded in a unified biological model of neurogenesis or neuroplasticity, relying instead on task- or heuristic-specific criteria.

Staying on the theory side cannot by itself prove algorithm effectiveness. Audio deepfake detection is an application that requires substantial feature understanding \cite{li2024audio, chen2020generalization}. While some small models have been shown to be effective \cite{pham2024deepfake}, and LLMs are also being explored for this task \cite{li2025dfallm}, practical deployment often raises additional challenges. In particular, selecting an appropriate network size is crucial: a model that is too small may underfit and fail to capture decisive features \cite{shahriar2026lightweight}, while a model that is too large may be inefficient and not generalise \cite{gu2025allm4add}. Dynamically adjusting networks with a neuroplasticity structure provides a flexible mechanism to maintain performance under computational constraints and evolving data distributions. 

\section{Methodology: Neuron-level dropin \& Plasticity}
\label{sec:illust}
Our method addresses the parameter bottleneck by selectively adding neurons and parameters to alleviate capacity limitations. Specifically, we propose dropin neuron addition on certain layers as a preliminary step to validate the effectiveness of neuron-level expansion. To avoid uncontrolled model growth while leveraging the benefits of dropin, we further introduce plasticity, which first adds neurons and subsequently prunes them to maintain a compact model size.

\subsection{Dropin algorithm}
\begin{figure}[ht]
    \centering
    \includegraphics[width=0.8\linewidth]{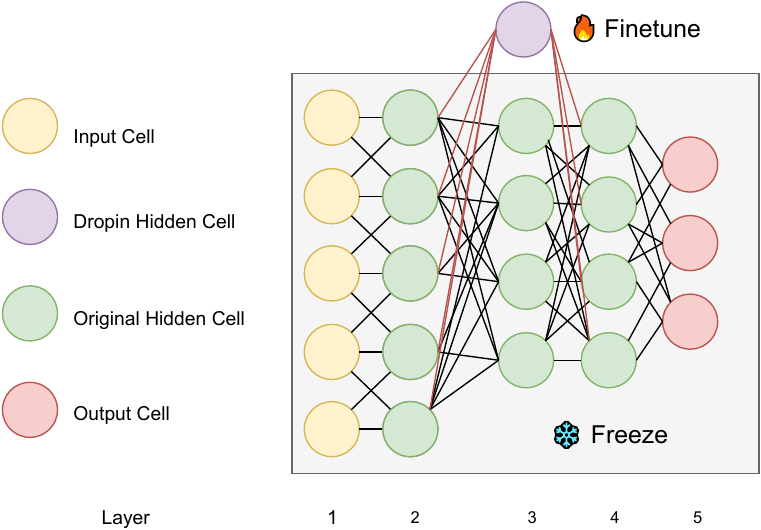}
    \caption{High-level dropin process: During the dropin phase, we load the original pretrained model weights and keep them frozen. Additional neurons are then introduced into randomly selected layers, and only the connection weights associated with these newly added neurons are trained.}
    \label{fig:dropin}
\end{figure}
To more closely emulate the neurogenesis observed in the mammalian brain, rather than extending entire layers, we introduce additional neurons selectively into specific layers. We add neurons to randomly selected layers and freeze the remainder of the network, training only the newly introduced neurons to examine their effects and behavior. 
This design choice is made as the complexity of determining where and when neurons are added in the mammalian brain is not easy to replicate in common state-of-the-art artificial neural networks.
%This design choice is motivated by the complexity of determining where and when neurons are added in the mammalian brain; 
%therefore, we begin with a general experimental setting to validate this mechanism in artificial neural networks. 
In addition to a a general experimental validation of the mechanism, we evaluate the effects of adding neurons at different layers from a statistical perspective. This approach operates at a more granular level and offers greater flexibility for the learning process, as individual neurons can be modulated directly by the algorithm. The overall framework is illustrated in Fig. \ref{fig:dropin}.

\begin{figure}
    \centering
    \includegraphics[width=0.9\linewidth]{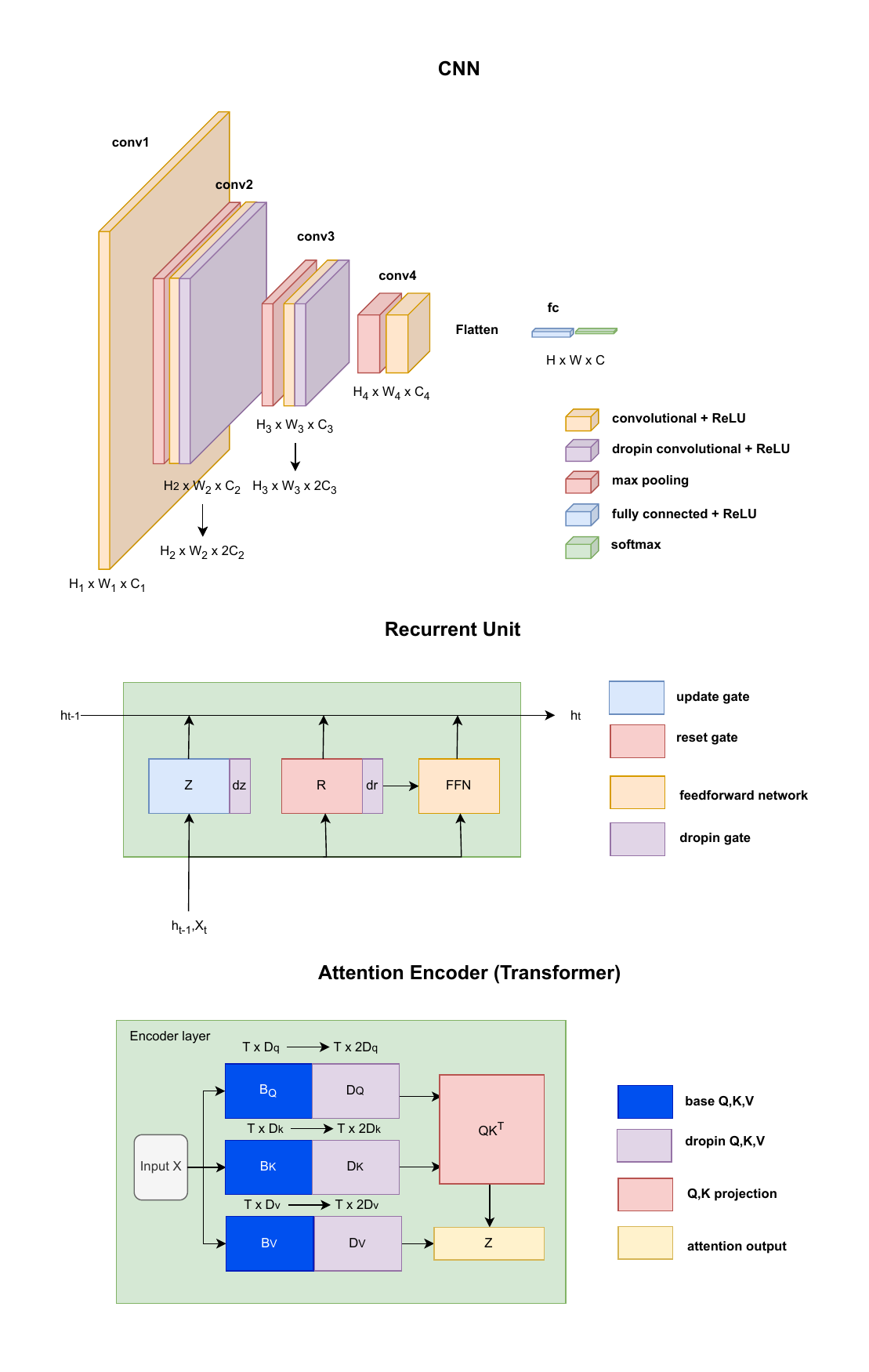}
    \caption{Convolutional layers, recurrent units, and attention encoders dropin process. These represent the specific dropin techniques adapted to different model architectures. From a mathematical perspective, this involves increasing the kernel size for CNNs, expanding the gate weight dimensions for GRUs, and enlarging the query, key, and value weight dimensions for attention mechanisms.}
    \label{fig:specific dropin}
\end{figure}

The above figure provides a simplified representation of a general neural network, in which each cell may represent any type of neuron and the connections correspond to the parameters of the model. As such, the proposed algorithm is not restricted to a specific architecture and can be applied to various network designs, including convolutional, encoder-based, and recurrent layers. More detailed illustrations of the implementation of our dropin mechanism for different architecture types are presented in Fig. \ref{fig:specific dropin}.

The dropin weights are concatenated as block matrices to facilitate matrix multiplication, enabling architectural adaptation at the granularity of individual neurons. Specifically, the channel dimension of certain layers is expanded, allowing additional neurons to be incorporated into the convolutional kernels for CNN. Likewise, the hidden dimensions of the update and reset gate weights in the GRU are enlarged, introducing more neurons within these gating mechanisms. Furthermore, in terms of attention encoder, the weight matrices used to generate the query, key, and value representations in attention mechanisms are expanded, permitting the inclusion of additional neurons. Importantly, the core computational mechanisms remain unchanged, for example, the convolution operations themselves are preserved, while the network capacity is increased through the addition of neurons. This demonstrates that the dropin approach is broadly applicable to a wide range of neural network architectures. For simplicity, we present a high-level version to illustrate the plasticity algorithm.

Additionally, this algorithm leverages pretrained models, which are now predominant in fine-tuning applications. In contrast to previous literature, which typically fine-tunes the entire network, our approach updates only the dropin neurons. This strategy considerably reduces training time while enhancing the model’s capability through a lightweight parameter increase, rather than relying on extensive stacking of layers. % such as ExHubert \cite{amiriparian2024exhubert}. 
Related work in continual learning has explored similar ideas, such as Low-Rank Adaptation (LoRA) \cite{hu2022lora}, which introduces additional low-rank matrices to acquire new information and update existing parameters. 
However, LoRA does not modify the network architecture to the same extent as our proposed drop-in mechanism. Instead, it introduces additional adapters that remain largely decoupled from the original architecture, making it less aligned with our biologically inspired motivation based on mammalian brain function.
%However, LoRA does not align with biologically inspired neural learning principles from mammalian brains at all, and does not modify the network architecture and is therefore less flexible. 
Additionally, LoRA is mainly suited for attention-based models, as it relies on updating a small number of parameters relative to the original weights. Therefore, its efficiency drops when many small layers lead to a large overall parameter count. In contrast, our method applies to any architecture by directly generating new neurons, independent of component parameterisation.

\subsection{Plasticity algorithm}
\begin{figure*}[t!]
    \centering
    \includegraphics[width=0.85\linewidth]{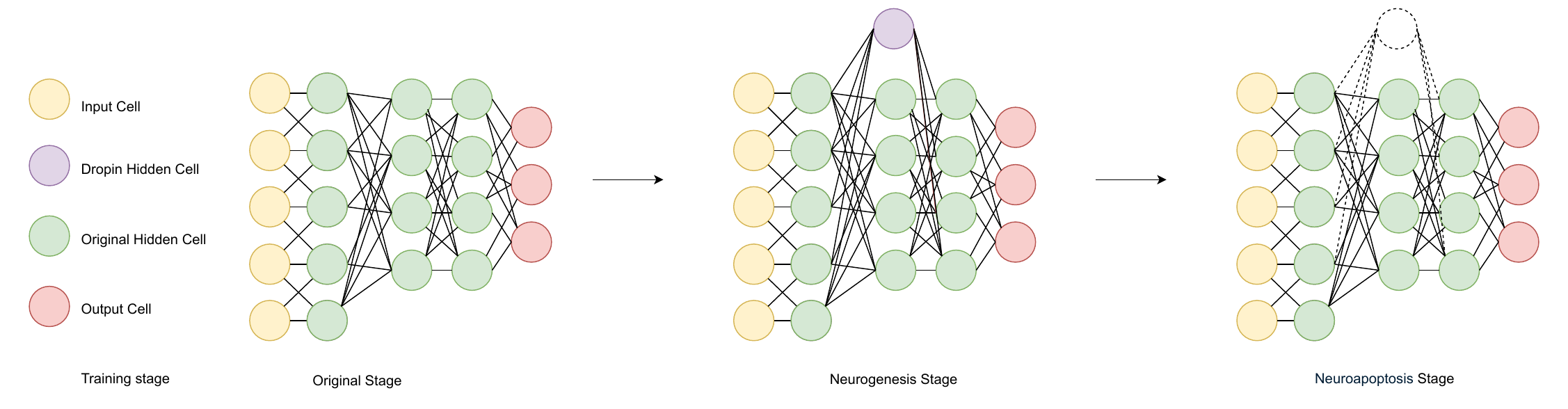}
    \caption{Plasticity process: The pipeline is divided into three stages. The first stage is identical to conventional training, where the model is optimised using the standard objective function. In the second stage, new neurons are dropped in, a process analogous to neurogenesis. The whole model is then retrained. After the newly introduced information has been assimilated and distributed across existing neurons. The third stage emulates neuroapoptosis by pruning the added neurons, followed by a final retraining.}
    \label{fig:plasticity}
\end{figure*}

Following the dropin process, the memory usage of the model increases. In contrast, the mammalian brain exhibits a natural mechanism of eliminating redundant or unused neurons to optimise efficiency. Building on the neuron-level granularity of the dropin mechanism, we take a further step by proposing a complementary algorithm, termed plasticity.
%BS: why now "Plasticity" when you always wrote "plasticity"? - please be coherent :)
The proposed pipeline is illustrated in Fig. \ref{fig:plasticity}.
This approach aims to push beyond the model’s capability limits on the target task. Accordingly, we design the entire training pipeline with an emphasis on maintaining the overall parameter count rather than optimising for time efficiency. The process begins with standard training of the model. \textbf{Unlike the controlled dropin experiments described above, where existing neurons are frozen to isolate the effects of newly added ones, in this stage we allow all parameters to remain trainable.} Our goal here is to investigate a complete biologically inspired setting from the neurogenesis to the neuroapotosis stage, in which network-wide interactions, analogous to biochemical processes in the brain, are permitted to emerge prior to the selective removal or deactivation of neurons.

The algorithm is divided into three stages. First, the model is trained for a number of epochs until its performance reaches a plateau. In the second stage, new neurons are introduced and the model continues training without reinitialising or freezing any previously learned parameters, such that all existing weights are updated jointly. This continuous training process enables the expanded architecture to adapt to the task and to reorganise its internal representations. As in the dropin procedure, the new neurons are added to randomly selected layers; for simplicity, the number of added neurons matches that of the original architecture. After this stage, the added neurons are pruned, and the full model undergoes a final phase of continued training, again with parameter initialisation from the previous training step. The algorithm is summarised in Algorithm \ref{al:plasticity}. 

This not only restores the original parameter size but also preserves the performance improvements gained during the intermediate expansion phase. We note that a promising direction for future investigation, once the mechanisms by which the brain prunes neurons are better understood, is to develop principled criteria for determining which neurons should be removed. At the current stage, however, we prune only the newly added neurons in order to control the number of trainable parameters and keep the overall model size constant. We contend that the plasticity algorithm could endow the model with a dynamic memory space, enabling it to retain valuable knowledge across all three stages while discarding irrelevant information, which forms the core intuition driving the model's breakthrough in performance. Each stage has been trained for equal epoches for a fair comparison.

% \begin{algorithm}[t]
% \caption{Neuron Dropin--Pruning Training}
% \label{al:plasticity}
% \begin{algorithmic}[1]
% \REQUIRE Network $\mathcal{N}(\theta)$ with layers $\{L_i\}_{i=1}^K$, dataset $\mathcal{D}$
% \ENSURE Final parameters $\theta^\ast$

% \STATE Sample layer indices:
% \[
% \mathcal{I} \sim \text{Uniform}(\{1,\dots,K\})
% \]

% \FOR{each $i \in \mathcal{I}$}
% \STATE Neuron expansion:
% \[
% L_i \leftarrow L_i \cup \Delta L_i,\quad
% |\Delta L_i| = |L_i|
% \]
% \ENDFOR

% \STATE Parameter initialization:
% \[
% \theta \leftarrow \theta \cup \theta_{\Delta},\quad
% \theta_{\Delta} \sim \mathcal{N}(0,\sigma^2)
% \]

% \STATE Joint optimization:
% \[
% \theta^{(1)} = \arg\min_{\theta}
% \mathbb{E}_{(x,y)\sim\mathcal{D}}
% \left[\mathcal{L}\big(f(x;\theta),y\big)\right]
% \]

% \STATE Neuron pruning:
% \[
% \theta^{(1)} \leftarrow \mathcal{P}\big(\theta^{(1)}\big)
% \]

% \STATE Final retraining:
% \[
% \theta^\ast = \arg\min_{\theta \subset \theta^{(1)}}
% \mathbb{E}_{(x,y)\sim\mathcal{D}}
% \left[\mathcal{L}\big(f(x;\theta),y\big)\right]
% \]

% \RETURN $\theta^\ast$
% \end{algorithmic}
% \end{algorithm}

\begin{algorithm}[t]
\caption{Neuroplasticity Training}
% \caption{Neuron Dropin--Pruning Training}
\label{al:plasticity}
\begin{algorithmic}[1]
\REQUIRE Network $\mathcal{N}(\theta)$ with layers $\{L_i\}_{i=1}^K$, dataset $\mathcal{D}$
\ENSURE Final parameters $\theta_\mathrm{pruned}^\ast$

\STATE Initial parameters:
\[
\theta_\mathrm{init}, \quad
\theta_\mathrm{init} \sim \mathcal{N}(0,\sigma^2)
\]

\STATE Initial training:
\[
\theta^{*}_\mathrm{init} = \arg\min_{\theta_\mathrm{init}}
\mathbb{E}_{(x,y)\sim\mathcal{D}}
\left[\mathcal{L}\big(f(x;\theta_\mathrm{init}),y\big)\right]
\]

\STATE Sample layer indices:
\[
\mathcal{I} \sim \text{Uniform}(\{1,\dots,K\})
\]

\FOR{each $i \in \mathcal{I}$}
\STATE Neuron expansion in layer $L^i$ with weights $\theta_\mathrm{init}^i$:
\[
\theta^i_\mathrm{dropin} \leftarrow \theta_\mathrm{init}^i \cup \theta_{\Delta}^i,\quad
\theta_{\Delta}^i \sim \mathcal{N}(0,\sigma^2), |\theta_{\Delta}^i| = |\theta_\mathrm{init}^i|
\]
% \[
% L_i \leftarrow L_i \cup \Delta L_i,\quad
% |\Delta L_i| = |L_i|
% \]
\ENDFOR

% \STATE Parameter initialization:
% \[
% \theta \leftarrow \theta \cup \theta_{\Delta},\quad
% \theta_{\Delta} \sim \mathcal{N}(0,\sigma^2)
% \]

\STATE Continued training with dropin neurons:
\[
\theta_\mathrm{dropin}^{*} = \arg\min_{\theta_\mathrm{dropin}}
\mathbb{E}_{(x,y)\sim\mathcal{D}}
\left[\mathcal{L}\big(f(x;\theta_\mathrm{dropin}),y\big)\right]
\]

\STATE Neuron pruning removes all added neurons:
\[
\theta_\mathrm{pruned} \leftarrow \mathcal{P}\big(\theta^{*}_\mathrm{dropin}\big), \quad |\theta_\mathrm{pruned}| = |\theta_\mathrm{init}|
\]
% \[
% \theta_\mathrm{dropout} \leftarrow \mathcal{P}\big(\theta^{*}_\mathrm{dropin}\big)
% \]

\STATE Final continued training:
\[
\theta_\mathrm{pruned}^* = \arg\min_{\theta_\mathrm{pruned}}
\mathbb{E}_{(x,y)\sim\mathcal{D}}
\left[\mathcal{L}\big(f(x;\theta_\mathrm{pruned}),y\big)\right]
\]

\RETURN $\theta_\mathrm{pruned}^\ast$
\end{algorithmic}
\end{algorithm}

\begin{table*}[!t]
\centering
\small
\resizebox{\linewidth}{!}{
\begin{tabular}{lllrrrr}
\toprule
\textbf{Dataset} & \textbf{Model} & \textbf{Training Strategy} &
\textbf{Test EER (\%)\,$\downarrow$} &
\textbf{Backward Time per step (ms)\,$\downarrow$} &
\textbf{Parameters (M)\,$\downarrow$} &
\textbf{Trainable Params (M)\,$\downarrow$} \\
\midrule

\multirow{18}{*}{ASV LA}
& \multirow{4}{*}{ResNet}
& baseline            & 14.69 & 5.46 & \textbf{11.17} & 11.17 \\
& & dropin unfrozen     & \textbf{7.60}  & 6.72 & 14.28 & 14.28 \\
& & dropin frozen (ours)& 11.98 & \textbf{3.49} & 14.28 & \textbf{1.47}  \\
& & plasticity (ours)   & 9.70  & /    & \textbf{11.17} & /     \\
\cline{2-7}

& \multirow{4}{*}{GRNN}
& baseline            & 19.84 & 343.35 & \textbf{0.06} & 0.06 \\
& & dropin unfrozen     & \textbf{16.64} & 477.79 & 0.10 & 0.10 \\
& & dropin frozen (ours)& 18.48 & \textbf{240.45} & 0.10 & \textbf{0.03} \\
& & plasticity (ours)   & 17.82 & /      & \textbf{0.06} & /    \\
\cline{2-7}

& \multirow{5}{*}{Wav2Vec 2.0}
& baseline            & 2.45 & 0.24 & \textbf{95.57} & 95.57 \\
& & dropin unfrozen     & 0.44 & 0.25 & 102.83 & 102.83 \\
& & LoRA                & 2.08 & /    & 95.57 & \textbf{0.29} \\
& & dropin frozen (ours)& 1.64 & \textbf{0.18} & 102.83 & 8.47 \\
& & plasticity (ours)   & \textbf{0.04} & /    & \textbf{95.57} & /    \\
\cline{2-7}

& \multirow{5}{*}{Wav2Vec 2.0 (Small)}
& baseline            & 11.06 & 1.15 & \textbf{11.25} & 11.25 \\
& & dropin unfrozen     & \textbf{8.44}  & 1.21 & 14.21 & 14.21 \\
& & LoRA                & 11.10 & /    & 11.25 & \textbf{0.09} \\
& & dropin frozen (ours)& 12.20 & \textbf{0.74} & 14.21 & 3.02 \\
& & plasticity (ours)   & 9.51  & /    & \textbf{11.25} & /    \\
\midrule
\multirow{18}{*}{ASV PA}
& \multirow{4}{*}{ResNet}
& baseline            & 14.08 & 5.82 & \textbf{11.17} & 11.17 \\
& & dropin unfrozen     & 9.14  & 6.84 & 14.28 & 14.28 \\
& & dropin frozen (ours)& \textbf{6.11}  & \textbf{3.79} & 14.28 & \textbf{1.47}  \\
& & plasticity (ours)   & 9.70  & /    & \textbf{11.17} & /     \\
\cline{2-7}

& \multirow{4}{*}{GRNN}
& baseline            & 19.13 & 355.88 & \textbf{0.06} & 0.06 \\
& & dropin unfrozen     & 18.68 & 494.30 & 0.10 & 0.10 \\
& & dropin frozen (ours)& \textbf{17.50} & \textbf{250.41} & 0.10 & \textbf{0.03} \\
& & plasticity (ours)   & 18.19 & /      & \textbf{0.06} & /    \\
\cline{2-7}

& \multirow{5}{*}{Wav2Vec 2.0}
& baseline            & 7.89 & 94.23 & \textbf{95.57} & 95.57 \\
& & dropin unfrozen     & 8.18 & 86.64 & 102.83 & 102.83 \\
& & LoRA                & 7.54 & /    & 95.57 & \textbf{0.29} \\
& & dropin frozen (ours)& 7.87 & \textbf{29.09} & 102.83 & 8.47 \\
& & plasticity (ours)   & \textbf{4.34} & /    & \textbf{95.57} & /    \\
\cline{2-7}

& \multirow{5}{*}{Wav2Vec 2.0 (Small)}
& baseline            & 11.98 & 21.94 & \textbf{11.25} & 11.25 \\
& & dropin unfrozen     & 14.72 & 25.25 & 14.21 & 14.21 \\
& & LoRA                & 16.52 & /    & 11.25 & \textbf{0.09} \\
& & dropin frozen (ours)& 11.97 & \textbf{10.27} & 14.21 & 3.02 \\
& & plasticity (ours)   & \textbf{10.75} & /    & \textbf{11.25} & /    \\

\midrule
\multirow{18}{*}{FoR}
& \multirow{4}{*}{ResNet}
& baseline            & 18.96 & 6.20 & \textbf{11.17} & 11.17 \\
& & dropin unfrozen     & \textbf{12.72} & 7.15 & 14.28 & 14.28 \\
& & dropin frozen (ours)& 13.70 & \textbf{3.97} & 14.28 & \textbf{1.47}  \\
& & plasticity (ours)   & 14.55 & /    & \textbf{11.17} & /     \\
\cline{2-7}

& \multirow{4}{*}{GRNN}
& baseline            & 17.65 & 368.81 & \textbf{0.06} & 0.06 \\
& & dropin unfrozen     & 16.27 & 506.93 & 0.10 & 0.10 \\
& & dropin frozen (ours)& 19.14 & \textbf{255.87} & 0.10 & \textbf{0.03} \\
& & plasticity (ours)   & \textbf{14.05} & /      & \textbf{0.06} & /    \\
\cline{2-7}

& \multirow{5}{*}{Wav2Vec 2.0}
& baseline            & 21.81 & 193.55 & \textbf{95.57} & 95.57 \\
& & dropin unfrozen                 & 21.13 & 202.77      & 102.83 & 102.83 \\
& & LoRA                & 16.67 & /      & 95.57 & \textbf{0.29} \\
& & dropin frozen (ours)                & 24.30 & \textbf{142.69}      & 102.83 & 8.47 \\
& & plasticity (ours)   & \textbf{12.94} & /      & \textbf{95.57} & /    \\
\cline{2-7}
& \multirow{5}{*}{Wav2Vec 2.0 (Small)}
& baseline            & 33.27 & 252.30 & \textbf{11.25} & 11.25 \\
& & dropin unfrozen                & 20.02 & 364.57      & 14.21 & 14.21 \\
& & LoRA                & 30.53 & /      & 11.25 & \textbf{0.09} \\
& & dropin frozen (ours)                & 19.82 & \textbf{209.59}      & 14.21 & 3.02 \\
& & plasticity (ours)   & \textbf{12.60} & /      & \textbf{11.25} & /    \\
\bottomrule
\end{tabular}
}
\caption{Comparison of different training strategies across ASV LA, PA, and FoR datasets.
Dropin unfrozen denotes an enlarged network trained conventionally, while baseline refers to the original model.
For plasticity, backward time and trainable parameters are not reported as they vary across stages.
The symbol ``/'' indicates unavailable or non-applicable measurements.}
\label{tab:main_results_all}
\end{table*}

\section{Experiments}
We evaluate our method on the ASVspoof 2019 Logical Access (LA),  physical access (PA) subset, and FoR, two widely used benchmarks for audio deepfake detection. Experiments are conducted on three representative models with distinct input modalities: ResNet18 (CNN, Mel spectrograms), GRNN (RNN, linear cepstral coefficients), and Wav2Vec 2.0 (attention-based, raw waveforms). This setup enables assessment of both architectural and feature diversity to verify the scalability of our pipeline. We fix the random seed to 42 for all random number generators (Python, NumPy, PyTorch) to ensure reproducibility. All experiments are run on an NVIDIA A6000 GPU, with other hyperparameters aligned to those in the original model papers \cite{he2015deepresiduallearningimage, baevski2020wav2vec20frameworkselfsupervised,gomez2019light}. We report the best-performing model for each setting. The maximum number of training epochs is determined through preliminary trials to ensure that each experimental configuration can converge within the allotted epochs. Model selection is performed based on validation set performance, and the selected model is subsequently evaluated on the test set. Wav2Vec 2.0 uses the pretrained model from Hugging Face\footnote{\url{https://huggingface.co/facebook/wav2vec2-base}}, while ResNet18 and GRNN are trained from scratch for all experimental settings including baseline. To examine parameter scaling, we also train a scratch Wav2Vec variant with a reduced 256-dimensional hidden size. Full hyperparameter details are provided in Github link \footnote{\url{https://github.com/ShawnNotFound/Dropin}}.

We apply the dropin and plasticity procedures once for each base model, introducing the same number of additional neurons as the original size of that layer. We compare the performance of five configurations: (i) the original baseline model, (ii) the dropin model without parameter freezing (representing an enlarged model capacity), (iii) LoRA-based fine-tuning (where applicable), (iv) our dropin method, and (v) our plasticity method. 

\section{Results}
In the following, we discuss the results of all models on the three datasets presented in Table \ref{tab:main_results_all}.
%The results of all models on three datasets are shown in Table \ref{tab:main_results_all}. And we list the findings as follows.
\begin{figure*}[h]
    \centering
    \includegraphics[width=0.9\linewidth]{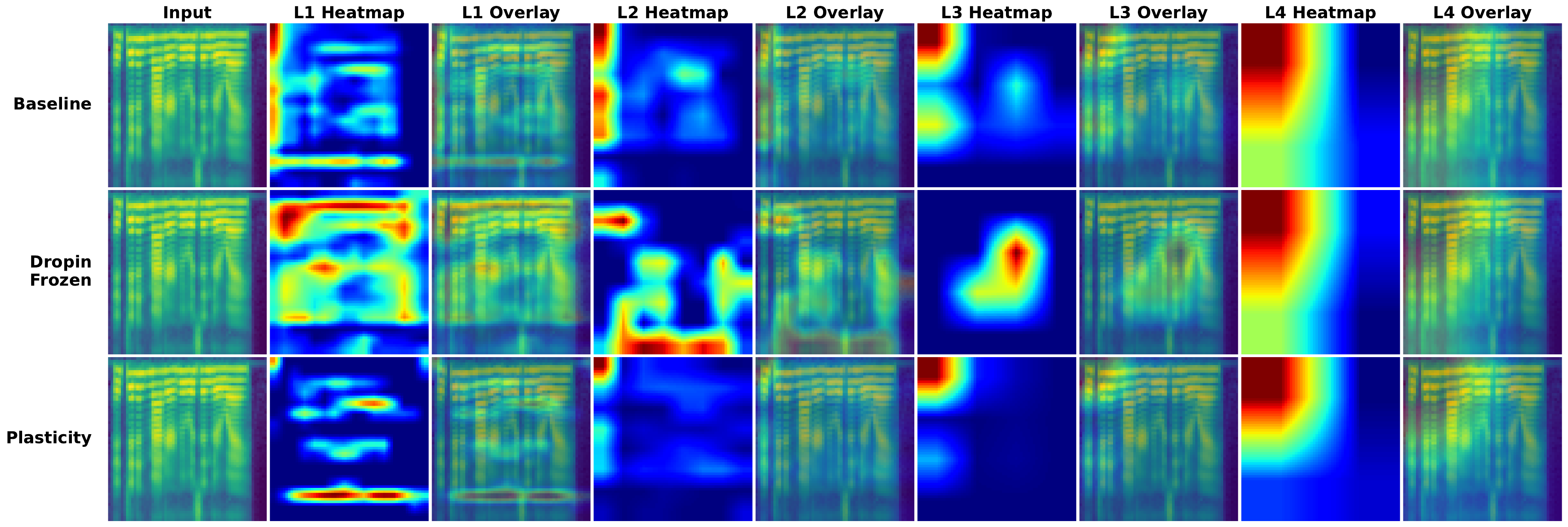}
    \caption{GradCAM results on Resnet for one case from ASVSpoof 2019 LA.}
    \label{fig:gradcam}
\end{figure*}

\begin{figure}[ht]
    \centering
    \includegraphics[width=0.85\linewidth]{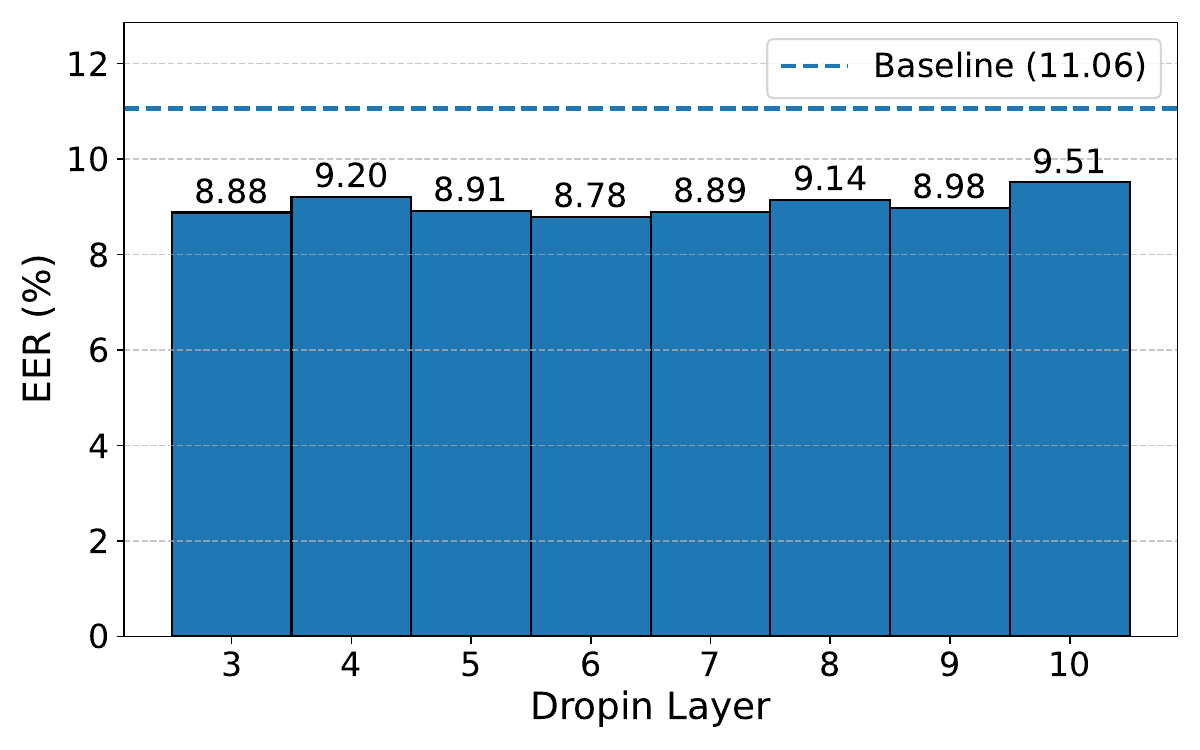}
    \caption{EER results for different dropin layers. ``3'' represents that we dropped in new neurons on the 3rd encoding layer in Wav2vec 2.0 small on ASVSpoof2019 LA.}
    \label{fig:eer_dropped_layers}
\end{figure}

\textbf{Accuracy improvement on plasticity:} The proposed dropin and plasticity strategies consistently outperform the baseline across multiple architectures and datasets in terms of Test EER. In particular, plasticity achieves the lowest EER in most settings, often by a large margin. For example, on the ASV LA dataset with Wav2Vec~2.0, plasticity reduces the EER from 2.45\% (baseline) to 0.04\%, substantially outperforming both dropin unfrozen (0.44\%) and dropin frozen (1.64\%). Similar improvements can be observed across ResNet and GRNN models on ASV LA and PA, demonstrating that the proposed mechanisms provide consistent accuracy gains over the original architectures.

\textbf{Efficiency improvement on dropin:} In terms of training efficiency, the dropin frozen strategy consistently achieves the lowest backward time among all compared methods. By freezing the original network and training only the newly introduced neurons, dropin frozen considerably reduces computational overhead while maintaining competitive detection performance. For instance, on ASV LA with ResNet, the backward time is reduced from 5.46,ms (baseline) to 3.49,ms, and on Wav2Vec~2.0, it is reduced from 0.24,ms to 0.18,ms. This efficiency advantage is also observed across PA and FoR datasets, indicating that dropin frozen is particularly suitable for scenarios where training speed is a primary concern.

\textbf{Plasticity achieves a superior trade-off between parameter size and accuracy:} When both accuracy and model size are considered jointly, plasticity emerges as the most favorable strategy. While dropin unfrozen occasionally achieves slightly lower EER, this improvement is primarily attributable to its substantially larger number of trainable parameters. In contrast, plasticity maintains the original model size and achieves comparable—or in many cases superior—performance. For example, on ASV PA with Wav2Vec~2.0, plasticity achieves an EER of 4.34\%, compared to 7.54\% for LoRA and 8.18\% for dropin unfrozen, without increasing the number of trainable parameters. These results suggest that plasticity provides a more balanced trade-off between accuracy and computational cost for model size, making it particularly attractive for practical deployment.

\textbf{Plasticity benefits more from larger models:} We observe that plasticity yields more pronounced performance gains in larger-capacity models compared to smaller ones. In particular, when comparing Wav2Vec~2.0 with ResNet, plasticity leads to substantially larger reductions in Test EER for Wav2Vec~2.0 across all evaluated datasets. For example, on ASV LA, plasticity reduces the EER of Wav2Vec~2.0 from 2.45\% to 0.04\%, whereas the improvement on ResNet is more moderate, decreasing from 14.69\% to 9.70\%. This trend is consistently observed on ASV PA and FoR, suggesting that larger models with higher representational capacity provide a more favorable substrate for plasticity-driven adaptation. These results indicate that plasticity is particularly effective when applied to high-capacity architectures, where additional neurons and dynamic adaptation can be more fully exploited.

\textbf{Comparative drawbacks of LoRA:} Finally, LoRA consistently underperforms compared to both dropin-based strategies and plasticity across all evaluated models and datasets. Although LoRA introduces only a small number of trainable parameters, its detection performance remains close to or worse than the baseline in most cases. For instance, on ASV LA with Wav2Vec~2.0, LoRA achieves an EER of 2.08\%, which is notably higher than plasticity (0.04\%) and dropin unfrozen (0.44\%). Similar trends are observed on PA and FoR datasets, indicating that parameter-efficient fine-tuning alone is insufficient for capturing the complex spoofing patterns addressed by our proposed methods but more mammal brain mimicing is needed.

We have also conducted \textbf{ablation studies} on effects of which layer to perform these mechanism and analyse explainability why the results performing better for our proposed method. Using the Wav2vec 2.0 small model with the dropin strategy as representatives, the results are presented in Fig. \ref{fig:eer_dropped_layers}.

It can be observed that dropping in neurons at different layers has a relatively small impact on overall performance. This suggests that neural networks are able to automatically leverage the added parameters to improve learning, regardless of the specific layer in which they are introduced. Importantly, this consistency indicates that our observed improvements are not due to randomness but represent a systematic effect. 
% These findings also imply that selecting a single layer based solely on mathematical criteria, such as gradients as previous literature did, is unlikely to provide meaningful guidance, hence we did not perform such comparisons. More broadly, they highlight that the mechanisms underlying neuron integration in the brain are likely more complex, motivating future experiments to better understand these processes and to develop more effective strategies for dropin and plasticity adaptation.
While some of the previous literature implies that selecting single layers based solely on mathematical criteria, such as gradients would be beneficial, we would argue that the selection process should rather be informed by processes in the mammalian brain, which are still to be understood well enough.

Lastly, we applied Grad-CAM \cite{selvaraju2017grad} to each experimental setting to examine whether attention patterns change during training. As a case study, we selected the ResNet model on a single example from the ASVspoof 2019 LA dataset. The resulting attention visualisations are presented in Fig.~\ref{fig:gradcam}.

It can be observed that in the baseline model, Grad-CAM activations are particularly prominent in the shallow layers (L1–L2), where focus is broadly distributed across edge regions. Although deeper layers show slightly more focused activations, the highlighted areas remain coarse. This suggests that the baseline relies on redundant or spurious features, resulting in limited feature selectivity and suboptimal hierarchical representations. However, under the dropin configuration, Grad-CAM maps become noticeably more compact from intermediate layers onwards. Compared to the baseline, the focus is more concentrated on many feature maps especially L1, which provides a potential explanation for the observed performance improvements. For the plasticity configuration, focus is even more selective across layers. Early layers exhibit relatively low activations, indicating enhanced feature selection capability, while deeper layers show highly focused and semantically aligned responses. By keeping parameters size the same, the model is able to adaptively reorganize feature representations, suppress non-informative responses, and achieve improved cross-layer consistency. Notably, the last layer’s attention of the three remains relatively similar, suggesting that there is still room for refinement in future work to further enhance focus and performance.

\section{Conclusion}
\label{sec:foot}
%BS: conclusion in past tense! I changed!
Our work investigated the use of the newly introduced dropin and plasticity algorithms, which can be applied to various model architectures. 
%BS: I reworded! 
We applied it in the field of audio deepfake detection to enhance both performance and efficiency in terms of computation and parameter usage. The model achieved SOTA performance on the ASVSpoof2019 LA subset. In scenarios with strict constraints on model size such as deployment on mobile devices or applications demanding high efficiency, including real-time update and detection tasks, our method presents a more practical and effective alternative. In future work, we aim to explore a biologically inspired plasticity algorithm that more closely mimics the behaviour of the mammalian brain. Specifically, this would involve structural pruning to remove neurons that are functionally redundant, a process that requires further rigorous definition. %BS: I changed all to British English, which was mostly chosen, but not always...

\bibliographystyle{IEEEtran}
\bibliography{reference}

@article{mckenzie2023inverse,
  title        = {Inverse Scaling: When Bigger Isn’t Better},
  author       = {McKenzie, Ian R. and Lyzhov, Alexander and Pieler, Michael and Parrish, Alicia and Mueller, Aaron and Prabhu, Ameya and McLean, Euan and Kirtland, Aaron and Ross, Alexis and Liu, Alisa and Gritsevskiy, Andrew and Wurgaft, Daniel and Kauffman, Derik and Recchia, Gabriel and Liu, Jiacheng and Cavanagh, Joe and Weiss, Max and Huang, Sicong and The Floating Droid and Tseng, Tom and Korbak, Tomasz and Shen, Xudong and Zhang, Yuhui and Zhou, Zhengping and Kim, Najoung and Bowman, Samuel R. and Perez, Ethan},
  journal      = {Transactions on Machine Learning Research},
  year         = {2023},
}

@article{chu2023qwen,
  title={Qwen-audio: Advancing universal audio understanding via unified large-scale audio-language models},
  author={Chu, Yunfei and Xu, Jin and Zhou, Xiaohuan and Yang, Qian and Zhang, Shiliang and Yan, Zhijie and Zhou, Chang and Zhou, Jingren},
  journal={arXiv preprint arXiv:2311.07919},
  year={2023}
}

@article{kaplan2020scaling,
  title={Scaling laws for neural language models},
  author={Kaplan, Jared and McCandlish, Sam and Henighan, Tom and Brown, Tom B and Chess, Benjamin and Child, Rewon and Gray, Scott and Radford, Alec and Wu, Jeffrey and Amodei, Dario},
  journal={arXiv preprint arXiv:2001.08361},
  year={2020}
}

@article{almutairi2022review,
  title={A review of modern audio deepfake detection methods: challenges and future directions},
  author={Almutairi, Zaynab and Elgibreen, Hebah},
  journal={Algorithms},
  volume={15},
  number={5},
  pages={155},
  year={2022},
  publisher={MDPI}
}

@inproceedings{he2015deepresiduallearningimage,
  title={Deep residual learning for image recognition},
  author={He, Kaiming and Zhang, Xiangyu and Ren, Shaoqing and Sun, Jian},
  booktitle={Proceedings of the IEEE conference on computer vision and pattern recognition},
  pages={770--778},
  year={2016}
}

@inproceedings{gomez2019light,
  title={A light convolutional GRU-RNN deep feature extractor for ASV spoofing detection},
  author={Gomez-Alanis, Alejandro and Peinado, Antonio M and Gonzalez, Jose A and Gomez, Angel M},
  booktitle={Proc. Interspeech},
  volume={2019},
  pages={1068--1072},
  year={2019}
}

@article{baevski2020wav2vec20frameworkselfsupervised,
  title={wav2vec 2.0: A framework for self-supervised learning of speech representations},
  author={Baevski, Alexei and Zhou, Yuhao and Mohamed, Abdelrahman and Auli, Michael},
  journal={Advances in neural information processing systems},
  volume={33},
  pages={12449--12460},
  year={2020}
}

@inproceedings{jung2022aasist,
  title={Aasist: Audio anti-spoofing using integrated spectro-temporal graph attention networks},
  author={Jung, Jee-weon and Heo, Hee-Soo and Tak, Hemlata and Shim, Hye-jin and Chung, Joon Son and Lee, Bong-Jin and Yu, Ha-Jin and Evans, Nicholas},
  booktitle={ICASSP 2022-2022 IEEE international conference on acoustics, speech and signal processing (ICASSP)},
  pages={6367--6371},
  year={2022},
  organization={IEEE}
}

@inproceedings{tak2021endtoendantispoofingrawnet2,
  title={End-to-end anti-spoofing with rawnet2},
  author={Tak, Hemlata and Patino, Jose and Todisco, Massimiliano and Nautsch, Andreas and Evans, Nicholas and Larcher, Anthony},
  booktitle={ICASSP 2021-2021 IEEE International Conference on Acoustics, Speech and Signal Processing (ICASSP)},
  pages={6369--6373},
  year={2021},
  organization={IEEE}
}

@article{wang2020asvspoof2019largescalepublic,
  title={ASVspoof 2019: A large-scale public database of synthesized, converted and replayed speech},
  author={Wang, Xin and Yamagishi, Junichi and Todisco, Massimiliano and Delgado, H{\'e}ctor and Nautsch, Andreas and Evans, Nicholas and Sahidullah, Md and Vestman, Ville and Kinnunen, Tomi and Lee, Kong Aik and others},
  journal={Computer Speech \& Language},
  volume={64},
  pages={101114},
  year={2020},
  publisher={Elsevier}
}

@misc{li2025neuroplasticityartificialintelligence,
      title={Neuroplasticity in Artificial Intelligence -- An Overview and Inspirations on Drop In \& Out Learning}, 
      author={Yupei Li and Manuel Milling and Björn W. Schuller},
      year={2025},
      eprint={2503.21419},
      archivePrefix={arXiv},
      primaryClass={cs.AI},
}

@inproceedings{amiriparian2024exhubert,
  title        = {ExHuBERT: Enhancing HuBERT Through Block Extension and Fine-Tuning on 37 Emotion Datasets},
  author       = {Amiriparian, Shahin and Packa{\'n}, Filip and Gerczuk, Maurice and Schuller, Bj{\"o}rn W.},
  booktitle    = {Proceedings of the Annual Conference of the International Speech Communication Association, INTERSPEECH},
  year         = {2024},
  pages        = {2635--2639},
  doi          = {10.21437/Interspeech.2024-280},
  publisher    = {International Speech Communication Association},
  address      = {Kos Island, Greece},
}

@inproceedings{reimao2019dataset,
  title={For: A dataset for synthetic speech detection},
  author={Reimao, Ricardo and Tzerpos, Vassilios},
  booktitle={2019 International Conference on Speech Technology and Human-Computer Dialogue (SpeD)},
  pages={1--10},
  year={2019},
  organization={IEEE}
}

@article{dohare2024loss,
  title={Loss of plasticity in deep continual learning},
  author={Dohare, Shibhansh and Hernandez-Garcia, J Fernando and Lan, Qingfeng and Rahman, Parash and Mahmood, A Rupam and Sutton, Richard S},
  journal={Nature},
  volume={632},
  number={8026},
  pages={768--774},
  year={2024},
  publisher={Nature Publishing Group UK London}
}

@inproceedings{selvaraju2017grad,
  title={Grad-cam: Visual explanations from deep networks via gradient-based localization},
  author={Selvaraju, Ramprasaath R and Cogswell, Michael and Das, Abhishek and Vedantam, Ramakrishna and Parikh, Devi and Batra, Dhruv},
  booktitle={Proceedings of the IEEE international conference on computer vision},
  pages={618--626},
  year={2017}
}

@article{schmidgall2024brain,
  title={Brain-inspired learning in artificial neural networks: a review},
  author={Schmidgall, Samuel and Ziaei, Rojin and Achterberg, Jascha and Kirsch, Louis and Hajiseyedrazi, S and Eshraghian, Jason},
  journal={APL Machine Learning},
  volume={2},
  number={2},
  year={2024},
  publisher={AIP Publishing}
}

@article{ravichandran2025unsupervised,
  title={Unsupervised representation learning with Hebbian synaptic and structural plasticity in brain-like feedforward neural networks},
  author={Ravichandran, Naresh and Lansner, Anders and Herman, Pawel},
  journal={Neurocomputing},
  volume={626},
  pages={129440},
  year={2025},
  publisher={Elsevier}
}

@article{rudroff2024neuroplasticity,
  title={Neuroplasticity meets artificial intelligence: A hippocampus-inspired approach to the stability--plasticity dilemma},
  author={Rudroff, Thorsten and Rainio, Oona and Klen, Riku},
  journal={Brain Sciences},
  volume={14},
  number={11},
  pages={1111},
  year={2024},
  publisher={MDPI}
}

@article{chen2025brain,
  title={Brain-Inspired Efficient Pruning: Exploiting Criticality in Spiking Neural Networks},
  author={Chen, Shuo and Liu, Zeshi and You, Haihang},
  journal={Concurrency and Computation: Practice and Experience},
  volume={37},
  number={27-28},
  pages={e70404},
  year={2025},
  publisher={Wiley Online Library}
}

@article{yousef2025synaptic,
  title={Synaptic plasticity-based regularizer for artificial neural networks},
  author={Yousef, Qais and Li, Pu},
  journal={Scientific Reports},
  volume={15},
  number={1},
  pages={14330},
  year={2025},
  publisher={Nature Publishing Group UK London}
}

@inproceedings{miconi2018differentiable,
  title={Differentiable plasticity: training plastic neural networks with backpropagation},
  author={Miconi, Thomas and Stanley, Kenneth and Clune, Jeff},
  booktitle={International Conference on Machine Learning},
  pages={3559--3568},
  year={2018},
  organization={PMLR}
}

@article{heidenreich2024transfer,
  title={Transfer learning of recurrent neural network-based plasticity models},
  author={Heidenreich, Julian N and Bonatti, Colin and Mohr, Dirk},
  journal={International Journal for Numerical Methods in Engineering},
  volume={125},
  number={1},
  pages={e7357},
  year={2024},
  publisher={Wiley Online Library}
}

@article{li2024audio,
  title={From Audio Deepfake Detection to AI-Generated Music Detection--A Pathway and Overview},
  author={Li, Yupei and Milling, Manuel and Specia, Lucia and Schuller, Bj{\"o}rn W},
  journal={arXiv preprint arXiv:2412.00571},
  year={2024}
}

@inproceedings{pham2024deepfake,
  title={Deepfake audio detection using spectrogram-based feature and ensemble of deep learning models},
  author={Pham, Lam and Lam, Phat and Nguyen, Truong and Nguyen, Huyen and Schindler, Alexander},
  booktitle={2024 IEEE 5th International Symposium on the Internet of Sounds (IS2)},
  pages={1--5},
  year={2024},
  organization={IEEE}
}

@article{li2025dfallm,
  title={DFALLM: Achieving Generalizable Multitask Deepfake Detection by Optimizing Audio LLM Components},
  author={Li, Yupei and Wang, Li and Wang, Yuxiang and Wang, Lei and Cai, Rizhao and Shi, Jie and Schuller, Bj{\"o}rn W and Wu, Zhizheng},
  journal={arXiv preprint arXiv:2512.08403},
  year={2025}
}

@article{shahriar2026lightweight,
  title={Lightweight Resolution-Aware Audio Deepfake Detection via Cross-Scale Attention and Consistency Learning},
  author={Shahriar, KA},
  journal={arXiv preprint arXiv:2601.06560},
  year={2026}
}

@inproceedings{gu2025allm4add,
  title={Allm4add: Unlocking the capabilities of audio large language models for audio deepfake detection},
  author={Gu, Hao and Yi, Jiangyan and Wang, Chenglong and Tao, Jianhua and Lian, Zheng and He, Jiayi and Ren, Yong and Chen, Yujie and Wen, Zhengqi},
  booktitle={Proceedings of the 33rd ACM International Conference on Multimedia},
  pages={11736--11745},
  year={2025}
}

@article{coscrato2020nn,
  title={The NN-Stacking: Feature weighted linear stacking through neural networks},
  author={Coscrato, Victor and de Almeida Inacio, Marco Henrique and Izbicki, Rafael},
  journal={Neurocomputing},
  volume={399},
  pages={141--152},
  year={2020},
  publisher={Elsevier}
}

@article{wozniak2020deep,
  title={Deep learning incorporating biologically inspired neural dynamics and in-memory computing},
  author={Wo{\'z}niak, Stanis{\l}aw and Pantazi, Angeliki and Bohnstingl, Thomas and Eleftheriou, Evangelos},
  journal={Nature Machine Intelligence},
  volume={2},
  number={6},
  pages={325--336},
  year={2020},
  publisher={Nature Publishing Group UK London}
}

@article{dellaferrera2021learning,
  title={Learning in Deep Neural Networks Using a Biologically Inspired Optimizer},
  author={Dellaferrera, Giorgia and Wozniak, Stanislaw and Indiveri, Giacomo and Pantazi, Angeliki and Eleftheriou, Evangelos},
  journal={arXiv preprint arXiv:2104.11604},
  year={2021}
}

@article{journe2022hebbian,
  title={Hebbian deep learning without feedback},
  author={Journ{\'e}, Adrien and Rodriguez, Hector Garcia and Guo, Qinghai and Moraitis, Timoleon},
  journal={arXiv preprint arXiv:2209.11883},
  year={2022}
}

@article{rusu2016progressive,
  title={Progressive neural networks},
  author={Rusu, Andrei A and Rabinowitz, Neil C and Desjardins, Guillaume and Soyer, Hubert and Kirkpatrick, James and Kavukcuoglu, Koray and Pascanu, Razvan and Hadsell, Raia},
  journal={arXiv preprint arXiv:1606.04671},
  year={2016}
}

@article{hu2022lora,
  title={Lora: Low-rank adaptation of large language models.},
  author={Hu, Edward J and Shen, Yelong and Wallis, Phillip and Allen-Zhu, Zeyuan and Li, Yuanzhi and Wang, Shean and Wang, Lu and Chen, Weizhu and others},
  journal={ICLR},
  year={2022}
}

@InProceedings{pmlr-v188-maile22a,
  title = 	 {When, where, and how to add new neurons to ANNs},
  author =       {Maile, Kaitlin and Rachelson, Emmanuel and Luga, Herv\'e and Wilson, Dennis George},
  booktitle = 	 {Proceedings of the First International Conference on Automated Machine Learning},
  pages = 	 {18/1--12},
  year = 	 {2022},
  editor = 	 {Guyon, Isabelle and Lindauer, Marius and van der Schaar, Mihaela and Hutter, Frank and Garnett, Roman},
  volume = 	 {188},
  series = 	 {Proceedings of Machine Learning Research},
  month = 	 {25--27 Jul},
  publisher =    {PMLR},

}

@inproceedings{evci2022gradmax,
  title={GradMax: Growing Neural Networks using Gradient Information},
  author={Evci, Utku and van Merri{\"e}nboer, Bart and Unterthiner, Thomas and Vladymyrov, Max and Pedregosa, Fabian},
  booktitle={International Conference on Learning Representations (ICLR)},
  year={2022},

}

@inproceedings{chen2020generalization,
  title={Generalization of Audio Deepfake Detection.},
  author={Chen, Tianxiang and Kumar, Avrosh and Nagarsheth, Parav and Sivaraman, Ganesh and Khoury, Elie},
  booktitle={Odyssey},
  pages={132--137},
  year={2020}
}

@article{eriksson2019dynamic,
  title={Dynamic network architectures for deep q-learning: Modelling neurogenesis in artificial intelligence},
  author={Eriksson, Pontus and Westlund Gotby, Love},
  year={2019}
}

@inproceedings{sowrirajan2024enhancing,
  title={Enhancing Neurofuzzy Plasticity: A Fusion of LSTM and Artificial Neurogenesis},
  author={Sowrirajan, Anirudh and Srinivasan, Pranav and Srinivasan, Sundari Avanthikaa and Devi, S Prasanna},
  booktitle={2024 International Conference on Emerging Techniques in Computational Intelligence (ICETCI)},
  pages={150--154},
  year={2024},
  organization={IEEE}
}
\end{document}